\documentclass[12pt,preprint]{emulateapj}

\newcommand{\etal}{{et~al.\null}}
\newcommand{\eg}{{e.g.,}}

\newcommand{\ie}{{i.e.,}}
\newcommand{\kms}{km~s$^{-1}$}
\newcommand{\Oline}{[\ion{O}{3}]~$\lambda$5007}

\slugcomment{}

\shorttitle{Planetary Nebula Spectroscopy}
\shortauthors{Herrmann \& Ciardullo}

\begin{document}

\title{Planetary Nebulae in Face-On Spiral Galaxies. \\ II. Planetary Nebula Spectroscopy}

\author{Kimberly A. Herrmann\altaffilmark{1,2,3} and Robin Ciardullo\altaffilmark{1,2}}
\affil{Department of Astronomy \& Astrophysics, The Pennsylvania State University \\ 525 Davey Lab, University Park, PA 16802}
\email{herrmann@lowell.edu, rbc@astro.psu.edu}

\altaffiltext{1}{Visiting Astronomer, Cerro Tololo Inter-American Observatory (CTIO). CTIO is operated by the Association of Universities for Research in Astronomy, Inc.\ (AURA) under cooperative agreement with the National Science Foundation (NSF).}

\altaffiltext{2}{Visiting Astronomer, Kitt Peak National Observatory, National Optical Astronomy Observatories (NOAO), which is operated by AURA, Inc.\ under contract to the NSF.  The WIYN Observatory is a joint facility of the University of Wisconsin-Madison, Indiana University, Yale University, and NOAO.}

\altaffiltext{3}{Current address: Lowell Observatory, 1400 West Mars Hill Road, Flagstaff, AZ, 86001}

\begin{abstract}
As the second step in our investigation of the mass-to-light ratio of spiral disks, we present the results of a spectroscopic survey of planetary nebulae (PNe) in five nearby, low-inclination galaxies: IC~342, M74 (NGC~628), M83 (NGC~5236), M94 (NGC~4736), and M101 (NGC~5457).  Using 50 setups of the WIYN/Hydra and Blanco/Hydra spectrographs, and 25 observations with the Hobby-Eberly Telescope's Medium Resolution Spectrograph, we determine the radial velocities of 99, 102, 162, 127, and 48 PNe, respectively, to a precision better than 15~\kms.   Although the main purpose of this data set is to facilitate dynamical mass measurements throughout the inner and outer disks of large spiral galaxies, our spectroscopy has other uses as well.  Here, we co-add these spectra to show that to first order, the [\ion{O}{3}] and Balmer line ratios of planetary nebulae vary little over the top $\sim$1.5~mag of the planetary nebula luminosity function.  The only obvious spectral change occurs with [\ion{N}{2}], which increases in strength as one proceeds down the luminosity function.  We also show that typical [\ion{O}{3}]-bright planetaries have $E(B-V) \sim 0.2$ of circumstellar extinction, and that this value is virtually independent of [\ion{O}{3}] luminosity.  We discuss the implications this has for understanding the population of PN progenitors. 

\end{abstract}


\section{INTRODUCTION}
Planetary Nebulae (PNe) are dying low-mass ($M \lesssim 8M_{\odot}$) stars whose ejected outer layers undergo ionization by the intense radiation from their central cores \citep[\eg][]{iau209}.  Their resulting spectra, which are dominated by the bright lines of [\ion{O}{3}] $\lambda\lambda 4959,5007$ and H$\beta$ in the blue, and H$\alpha$ and [\ion{N}{2}] $\lambda\lambda 6548,6584$ in the red, are ideally suited for radial velocity programs.  Moreover, because PNe are bright, plentiful, distinctive, and representative of an older stellar population, they are the objects of choice for a host of kinematic problems, ranging from the measurement of dark matter in elliptical galaxies \citep{PNS, deLorenzi} to the study of galaxy interactions \citep{M51, CenA}.  

PNe have been heavily used to study the kinematics of early-type galaxies \citep[][and references therein] {sw06,PNearly} but similar studies in late-type spirals have been lacking, due mostly to the problems associated with PN identification.  In star-forming systems, \ion{H}{2} regions far outnumber bright planetary nebulae, so unless one works in the Local Group where the contaminating objects can be spatially resolved \citep[\eg][]{M33,M31PNS}, or in the halos of edge-on systems \citep[\eg][]{p7,p10}, extreme care is needed to discriminate between the two classes of objects. In fact, PN-based kinematic studies have been performed in the disks of only a few late-type systems: the SMC \citep[44 objects;][]{dlfw85}, the LMC \citep[110 objects;][]{mdfw88,vmd92}, M94 \citep[67 objects;][]{M94PNS}, M33 \citep[140 objects;][]{M33}, and M31 \citep[$>$2000 objects;][]{M31PNS}.

\footnotetext[4]{The Hobby-Eberly Telescope (HET) is a joint project of the University of Texas at Austin, the Pennsylvania State University, Stanford University, Ludwig-Maximillians-Universit\"at M\"unchen, and Georg-August-Universit\"at G\"ottingen.  The HET is named in honor of its principal benefactors, William P. Hobby and Robert E. Eberly.}

\begin{deluxetable*}{llccccccl}
\tabletypesize{\scriptsize}
\tablecaption{Target Galaxies\label{tabBasic}}
\tablewidth{0pt}
\tablehead{ &&&\colhead{$v_{\odot}$\tablenotemark{a}}
&\colhead{Distance\tablenotemark{b}} &\colhead{Survey}&&&\\
\colhead{Galaxy} &\colhead{Type} &\colhead{Size\tablenotemark{a}}
&\colhead{(\kms)} &\colhead{(Mpc)} &\colhead{Region} &\colhead{P.A.} 
&\colhead{$i$} &\colhead{P.A. and $i$ Reference}
}
\startdata
IC~342  &Scd &$21\farcm4$ & 34 &$3.5 \pm 0.3$ &$4\farcm 8$ &$39^\circ$ &25$^\circ$  &\citet{N80}  \\
M74     &Sc  &$10\farcm5$ &656 &$8.6 \pm 0.3$ &$4\farcm 8$ &$25^\circ$ &6.5$^\circ$ &\citet{KB92} \\
M83     &SBc &$12\farcm9$ &516 &$4.8 \pm 0.1$ &$18\arcmin$ &$46^\circ$ &24$^\circ$  &\citet{L+04}  \\
M94     &Sab &$11\farcm2$ &310 &$4.4^{+0.1}_{-0.2}$ &$5\farcm 8$ &$115^\circ$ &35$^\circ$  &\citet{MC96}  \\
M101    &Scd &$28\farcm8$ &241 &$7.7 \pm 0.5$ &$8\arcmin$  &$35^\circ$ &17$^\circ$  &\citet{ZEH90}  \\
\enddata
\tablenotetext{a}{From RC3}
\tablenotetext{b}{From Paper~I except for M101, which is from \citet{M101PNe}}
\end{deluxetable*}

In Paper~I \citep{thesis1}, we presented the results of narrow-band [\ion{O}{3}] and H$\alpha$ surveys for PNe in six nearby, low-inclination galaxies:  IC~342, M74 (NGC~628), M83 (NGC~5236), M94 (NGC~4736), NGC~5068, and NGC~6946.  Here, we present follow-up PN spectroscopy in the first four of these systems (each with $>$140 PNe), plus M101 (NGC~5457), a galaxy with prior PN identifications from \citet{M101PNe}.  In \S2 we describe our observations with the Hydra multi-fiber spectrographs of the WIYN and Blanco telescopes, and detail our supplemental observations with the Medium Resolution Spectrograph (MRS) of the Hobby-Eberly Telescope (HET)\footnotemark[4].  In \S3, we outline the reduction procedures required to extract measurable spectra from these instruments.  We explore the precision of our radial velocities in \S4 using a series of tests, including a $\chi^2$ analysis of spectra taken with different setups on different nights, and an external comparison with the results of counter-dispersed imaging \citep{M94PNS}. In \S5, we describe our efforts to remove contaminating objects, such as \ion{H}{2} regions and background emission galaxies, from our sample.  In \S6, we present our final PN velocities and uncertainties; these measurements serve as the basis of a kinematic analysis of the systems' inner and outer disks \citep[][(Paper~III)]{letter,thesis3}.  We also co-add the data to produce a series of ``mean'' spectra for our extragalactic planetaries, and explore the excitation of these spectra as a function of [\ion{O}{3}] $\lambda 5007$ absolute magnitude.  Finally, in \S7, we discuss the origins of our PN progenitors and test for population homogeneity using circumstellar extinction.  Our conclusions are in \S8.

\section{OBSERVATIONS}
In Paper~I, we surveyed the PN populations of six large ($r > 7\arcmin$), nearby ($D < 10$~Mpc), low-inclination ($i < 35^\circ$) spiral galaxies with \Oline\ and H$\alpha$ imaging.  From this sample, four PN systems were deemed large enough for kinematic follow-up:  those of IC~342, M74, M83, and M94.  In addition, we also targeted the PN system of M101, a galaxy which had been previously surveyed by \citet{M101PNe}.   A description of these systems is given in Table~\ref{tabBasic}.

Our goal was to obtain precise ($\lesssim 15$~\kms) velocities for as many of our previously identified planetary nebula candidates as possible.  To do this, we used the Hydra multi-fiber spectrographs on the WIYN and Blanco telescopes, supplemented with spectra from the Medium Resolution Spectrograph (MRS) of the Hobby-Eberly Telescope (HET).  For the Hydra runs, our strategy was to maximize the number of PNe targeted, while minimizing our velocity errors, all under the constraint imposed by the fiber positioners (\ie\ the requirement of keeping a minimum fiber separation of $37\arcsec$ on WIYN and $25\arcsec$ on Blanco).  To work around this constraint, which typically limited us to observing $\lesssim 40$~PNe per setup, we performed quick-look data reductions immediately after each observation.  By assessing the quality of each spectrum in real time, we rapidly identified those PNe needing additional data and gave them a higher priority in the next night's fiber assignments.  In this way, we not only maximized the number of PNe with high-precision velocities, but also controlled the systematic errors associated with observations through different fibers and different Hydra configurations.

Our Hydra spectroscopy in the north was performed with the 3.5-m WIYN telescope at Kitt Peak during 6 separate runs between March 2003 and November 2007.  The first of these runs targeted the PNe of M101 with $2\arcsec$ red-sensitive fibers and a 600 lines mm$^{-1}$ grating blazed at $10\fdg 1$ in first order, producing spectra between 4500 and 7000 \AA, at a dispersion of 1.4 \AA\ pixel$^{-1}$ with $\sim$4.6~\AA\ (275 \kms) resolution.  Our subsequent observations used the same fiber bundle, but with a new 740 lines~mm$^{-1}$ Volume Phase Holographic (VPH) grating, designed to optimize throughput near 4990~\AA\null.  These spectra covered the wavelength range from 4400~\AA\ to 5500 \AA, with higher dispersion (0.5 \AA\ pixel$^{-1}$), improved resolution (1.4 \AA, or 84 \kms) and, most importantly, greater efficiency.  Each Hydra setup was observed for 3 hours, typically using a series of four 45~min exposures.

For our southern (M83) observations, we used the version of Hydra on the CTIO 4-m Blanco telescope, with an atmospheric dispersion corrector, $2\arcsec$ STU fibers, and a 632~lines~mm$^{-1}$ grating blazed at $10\fdg 8$ in first order.  This instrument yielded spectra with a resolution of 3.3~\AA\ (198~\kms) and a dispersion of 0.59~\AA~pixel$^{-1}$ over the wavelength range between 4500 and 6900~\AA\null.  Again each Hydra setup consisted of a series of 45~min exposures totalling 3 hours.  However, because of the instrument's larger number of fibers (138 versus 86) and smaller minimum fiber separation ($25\arcsec$ versus $37\arcsec$), and because our photometric survey of M83 encompassed a much wider field-of-view than those for our other galaxies, each setup was able to target $\sim$70~PNe at once, rather than just $\sim$40.

Finally, to supplement our Hydra observations, we targeted some of the M101 PNe with the Medium Resolution Spectrograph of the queue-scheduled Hobby-Eberly Telescope.  This dual-beam, bench-mounted instrument has a 79~lines~mm$^{-1}$ echelle grating, a 220~lines~mm$^{-1}$ cross-disperser, and a single $2\arcsec$ red-sensitive fiber which delivers data from 4400 to 6200~\AA\ in the blue, and 6300 to 10,000~\AA\ in the red.  Although this instrument could only target objects within $50\arcsec$ of an $V < 17$ offset star, the MRS' high dispersion (0.14~\AA~pixel$^{-1}$ with 1.1~\AA\ resolution) coupled with the HET's large aperture produced precise velocities with relatively short ($\sim$20~min) exposures.  A log of all our observations appears in Table~\ref{tabObs}.

\begin{deluxetable*}{lllccccl}
\tabletypesize{\scriptsize}
\tablecaption{Observing Log\label{tabObs}}
\tablewidth{0pt}
\tablehead{
               &\colhead{Observing} &Telescope &\colhead{Number} &
\colhead{PNe} &\colhead{Sky Fibers}  &\colhead{Exposure} & \colhead{Sky} \\
\colhead{Galaxy} &\colhead{Dates} &\& Grating &\colhead{of Setups} 
&\colhead{per Setup} &\colhead{per Setup} &\colhead{Time (min)} 
&\colhead{Conditions} 
}
\startdata
IC 342 &2006 Nov 19-22 &WIYN/VPH  &5  &31-35  &6-7  &$4 \times 45$ &phot \\
IC 342 &2007 Mar 13-18 &WIYN/VPH  &4  &31-34  &6-7  &$4 \times 45$ &phot-spec\\
IC 342 &2007 Nov 10-12 &WIYN/VPH  &2  &35-36  &10   &$4 \times 45$ &cloudy \\
\\
M74  &2006 Oct 13    &WIYN/VPH  &1  &31     &5    &$4 \times 45$ &cloudy \\
M74  &2006 Nov 19-22 &WIYN/VPH  &7  &26-32  &5-9  &$4 \times 45$ &phot \\
M74  &2007 Nov 10-12&WIYN/VPH  &6  &29-32  &9-11 &$4 \times 45$ &cloudy\\
\\
M83  &2005 May 30-Jun 2 &Blanco/632@10.8 &8 &66-73 &10-50 &$4 \times 45$ &phot\\
\\
M94  &2006 Mar 2-5   &WIYN/VPH  &5  &25-28  &5-6  &$4 \times 45$ &phot-spec\\
M94  &2007 Mar 13-18 &WIYN/VPH  &6  &27-29  &5-7  &$4 \times 45$ &phot-spec\\
\\
M101 &2003 Mar 24-25 &WIYN/600@10.1 &4 &31-35 &12-23 &$6 \times 30$ &phot-spec\\
M101 &2005 Mar - 2006 Mar &HET/MRS  &25 &1    &0     &15 - 35       &phot-spec\\
M101 &2006 Mar 2-5   &WIYN/VPH    &2  &34     &4-5   &$4 \times 45$ &phot-spec\\
\enddata
\end{deluxetable*}

\section{SPECTRAL REDUCTION}

\footnotetext[5]{IRAF is distributed by NOAO, which are operated by AURA, Inc., under cooperative agreement with the NSF.}

\subsection{Hydra spectra}
Our data were reduced using the routines of IRAF\footnotemark[5] \citep{v98}.  To reduce the Hydra data, we began with the tasks within the {\tt ccdred} package:  the data were trimmed and bias-subtracted via {\tt ccdproc}, the dome flats (typically three per setup) were combined using {\tt flatcombine}, and the comparison arcs (CuAr at WIYN and Penray HeNeArXe at Blanco) which bracketed the target exposures were combined via {\tt imcombine}.  Next, {\tt dohydra} within the {\tt hydra} package was used to reduce the spectra, with the averaged dome flats serving to define the extraction apertures, and the averaged comparison arcs providing the wavelength calibration to a precision better than 0.03~\AA\ for WIYN+600@10.1, 0.02~\AA\ for Blanco, and 0.013~\AA\ for WIYN+VPH\null.  Finally, the individual spectra were re-sampled onto a log wavelength scale to facilitate the co-addition of data taken at different times of the year.

We note that the wavelength calibration of the CTIO data required some extra attention.  The Penray lamp's emission lines in the blue are $\sim$3 orders of magnitude weaker than its lines in the red.  Since very short and very long comparison arcs were only taken on the first night of the run, we used a spliced version of the arcs (\ie\ with a long exposure in the blue and a short exposure in the red) as the master comparison for all four nights' data.  To check for possible setup-to-setup variations, the inferred wavelengths of five strong, well-defined emission lines (\ion{Xe}{1} $\lambda 4671$, \ion{He}{1} $\lambda 5016$, \ion{Ne}{1} $\lambda 5401$, \ion{Ne}{1} $\lambda 6533$, and \ion{Ne}{1} $\lambda 6599$) on each individual exposure were compared to their wavelengths on the master arc.  For the first 6 setups of the run, this test showed no significant variations other than small zero point shifts.  However, for our last night's observations, the wavelengths of \ion{Ne}{1} $\lambda 6533$ and \ion{Ne}{1} $\lambda 6599$ were offset $\sim$10~\kms\ to the blue with respect to the other lines.  To correct for this shift, the H$\alpha$ velocities measured from the last two setups were incremented by this small amount.

After extracting each spectrum, the PNe were sky subtracted using data acquired through several blank-field fibers.  For the WIYN+VPH spectra, this step was straightforward:  since no bright sky lines fell within the wavelength range of the instrument, we simply used {\tt scombine} to combine the extracted spectra from the multiple exposures and {\tt skysub} for the subtraction.  For the WIYN+600@10.1 and Blanco spectra, which had wider wavelength coverage, we used {\tt scombine} as before, then used {\tt skytweak} to align the spectra before subtracting.  We note our observations were all taken during dark time and that no bright sky lines exist near any of the emission features of interest.  Consequently, the details of this step do not effect our final results.

Finally, each PN spectrum was shifted into the barycentric rest frame using the IRAF task {\tt dopcor}.  These velocity corrections were especially important for IC~342, a $\beta = 46^\circ$ object whose data were collected at different times of the year, but all our spectra were shifted, even if the correction was less than 1 \kms.  Once in the barycentric frame, the data from the multiple setups were co-added to create a final summed spectrum for each PN.

\subsection{MRS spectra}
Our echelle data from the Hobby-Eberly Telescope's Medium Resolution Spectrograph were reduced using an automated pipeline designed by K.A.H. to take advantage of the instrument's long-term stability.  First, the data from the blue and red sides of the spectrograph were trimmed and bias-subtracted with {\tt ccdproc}.  Next, as with Hydra, the spectra were extracted and flatfielded using the night's dome flats, and wavelength calibrated using ThAr comparison arcs.  We note that the latter step was usually performed via the echelle task {\tt ecreidentify} and an initial wavelength solution stored in the pipeline's database.  (The existence of this database also allowed us to test for time-dependent systematic errors in the wavelength calibration.)  Finally, the spectra were re-sampled onto a log-wavelength scale, shifted into the barycentric frame, and almost always co-added with spectra taken with WIYN+Hydra.  No sky subtraction was performed on these single-fiber observations.  Again, since the data were taken during dark time and our targeted spectral features are far from any sky line, this omission in no way changed our results.  

\section{MEASURING VELOCITIES AND UNCERTAINTIES}
\begin{figure}
\epsscale{1.1}
\plotone{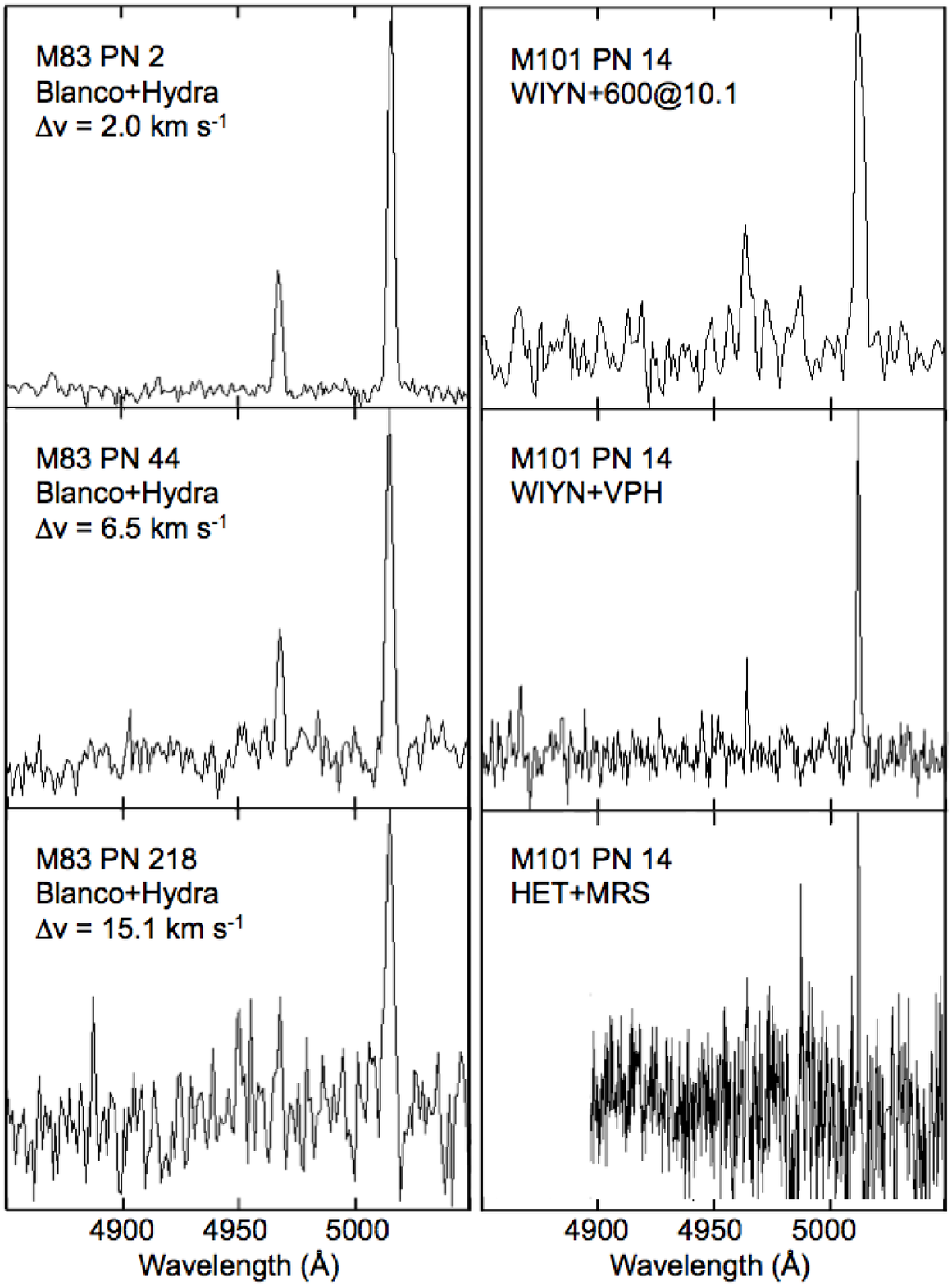}
\caption{Sample spectra from each of our four instrumental configurations in the wavelength range between 4850 and 5050 \AA.  Included are three spectra from the Blanco telescope, illustrating the relationship between data quality and velocity uncertainty ($\Delta v$).  Note that virtually all velocity information is contained in the \Oline; the flux in [\ion{O}{3}] $\lambda 4959$ is $\sim$3~times weaker, and H$\beta$ is barely visible. \label{spectra} }
\end{figure}

Figure~\ref{spectra} gives sample spectra from each of our instrument configurations and illustrates the varying quality of our data.  Although a number of lines are present, the brightest feature, by far, is always the 5007~\AA\ emission from doubly-ionized oxygen.  We therefore determined our PN velocities (and velocity uncertainties) solely from this line, via the line-fitting routines of {\tt emsao} within IRAF's {\tt rvsao} package \citep{rvsao}.  Weaker lines, such as [\ion{O}{3}] $\lambda 4959$, H$\alpha$, and H$\beta$ were also measured, but since the precision of our line centroiding went almost linearly with counts, these additional features did not significantly improve the accuracy of our measurements.  They were, however, useful for exploring the systematics of the PN population (see \S7). 

In order to use PNe as kinematic probes in Paper~III, it is imperative that the uncertainty associated with each velocity measurement be known.  To investigate this number, we began with a straightforward internal consistency check.  Since most of our spectra were acquired using a sequence of four 45-minute exposures, we simply divided the datasets in two and compared the velocities found from the co-addition of the first two exposures with those derived from the co-addition of the last two frames.  Obviously, since each spectral subset contained only half the signal of the whole, only the brightest PNe could be analyzed in this way.  Consequently  we restricted this analysis to the PNe of M83 and M94.  Moreover, since the combination of two exposures was not always sufficient for eliminating cosmic ray hits, a number of objects were found to have wildly discrepant results.  Nevertheless, this comparison demonstrated that any systematic drift over the 3~hr period of observation was minimal.  An analysis of the 119 PN observations in M94 showed just random scatter and in 108 of the cases, the two measurements were within the $1\,\sigma$ internal uncertainty.  Similarly, our comparison of 314~PN velocity pairs in M83 yielded only 63 objects with more than $1\,\sigma$ errors, and again, found no evidence for any time-dependent velocity shift.  Not only does this confirm the validity of our 3 hour co-additions, but it also suggests that the internal uncertainties returned by {\tt emsao} are reasonable.

We next tested our data for systematic errors associated with the independent fiber setups.  To do this, we took advantage of the fact that many of our PNe were targeted multiple times using different fibers and different wavelength calibrations.  We began by intercomparing all our PN spectra, and identifying those objects whose [\ion{O}{3}] line flux differed drastically from one setup to the next.  In these cases, where co-addition would have only degraded the signal, the lower signal-to-noise observation was dropped from the analysis.  We then combined the data to create a single, summed spectrum for each planetary, and with the aid of {\tt emsao}, we measured each object's velocity and velocity uncertainty.  Next, we compared these summed velocities to the velocities found with the spectra of the individual fiber-setups.  By combining the results from all the planetaries observed using multiple setups, we were able to infer the mean velocity offset of each fiber configuration.

\begin{figure*}
\epsscale{1.1}
\plotone{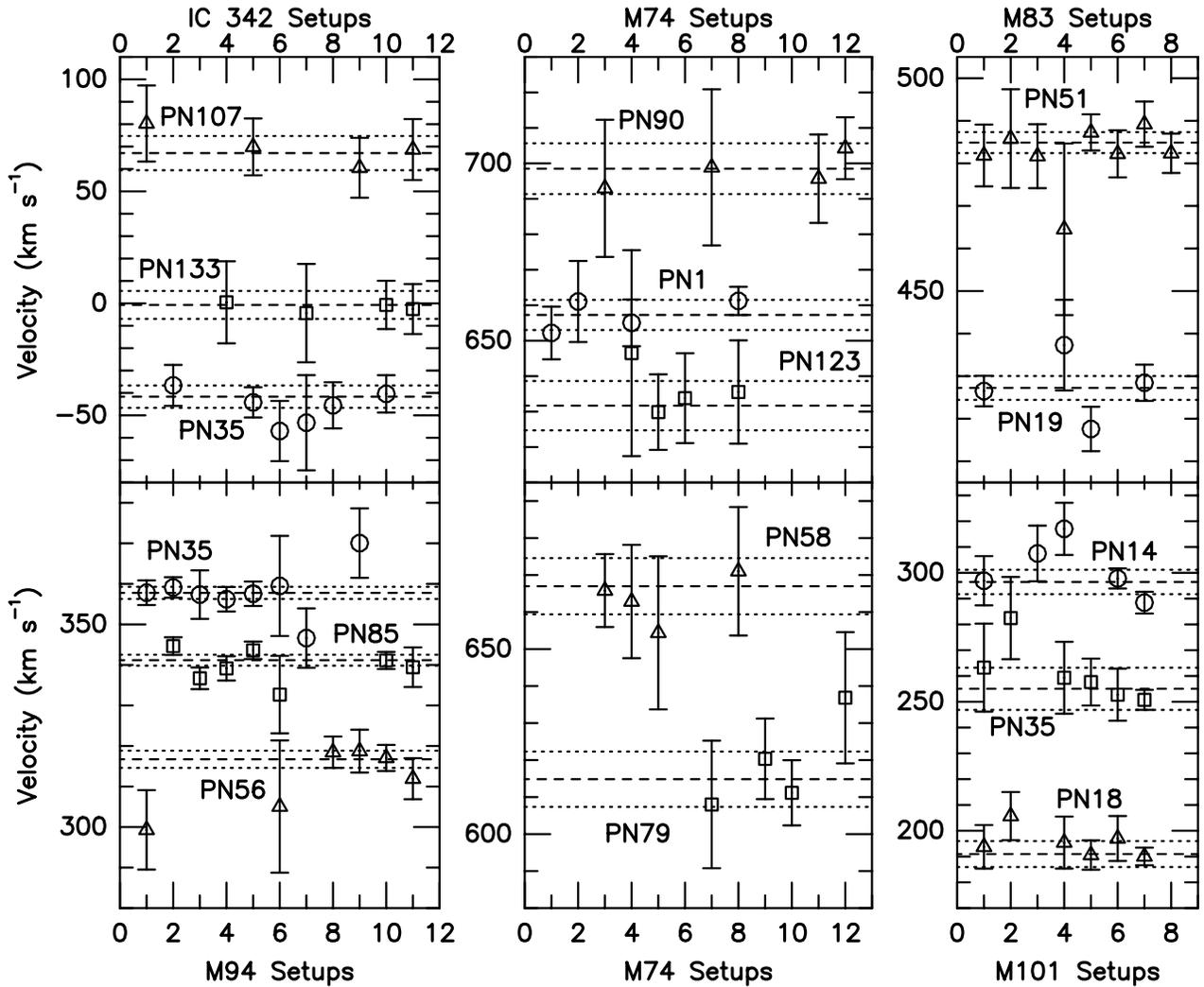}
\caption{Velocities of objects observed multiple times as a function of setup number.  The dashed line indicates the final velocity derived from the summation of all exposures;  the dotted lines on either side of this mean indicate the $1 \, \sigma$ uncertainty of the final value.  Note the generally good agreement: measurements that are discrepant usually have large uncertainties.  M94's setup 6 was affected by clouds, which explains its significantly larger error bars. \label{mult_obs} }
\end{figure*}

The results from this analysis are given in Table~\ref{tabBetSetups}.  As can be seen, there is scant evidence of any systematic velocity shift associated with the individual setups.  Of the 51 configurations used in this program (where we group all the M101 MRS spectra as being in the galaxy's Setup~7), over half have offsets within one standard deviation of the mean, and $\sim$80\% have offsets that agree to within $2 \,\sigma$.  This means that, at most, the systematic error associated with each individual setup is $\sim$1.6~\kms.  (Such an error would increase the number of $2 \, \sigma$ agreements to 96\%, the number appropriate for a Gaussian distribution.)  In addition, in only four cases is the amplitude of the systematic shift observed to be greater than 5~\kms:  Setup~9 in IC~342 ($+9.2 \pm 4.3$~\kms), Setup~2 in M83 ($+5.3 \pm 2.0$~\kms), Setup~1 in M101 ($+6.1 \pm 2.4$~\kms), and Setup~5 in M101 ($+8.2 \pm 4.0$~\kms).  In none of these cases does the offset even approach $3 \, \sigma$; this consistency again confirms that any systematic shift between individual fiber setups is minimal.   An illustration of our measurement stability is shown in Figure~\ref{mult_obs}, where the velocities and velocity uncertainties of several of the most observed PNe are plotted. 

Since the systematic errors in our velocity measurements are minimal, we can use our data to test whether the internal errors computed by {\tt emsao} represent the true uncertainties of our measurements.  To do this, we performed a pairwise comparison of all the PNe observed with multiple setups, using the 
$\chi^2$ statistic
\begin{equation}
\chi^2 = \sum_{i\neq j} {\left(v_i - v_j\right)^2 \over \sigma_i^2 + \sigma_j^2}\label{chi2}.
\end{equation}
In the equation, $v_i$ and $v_j$ represent the independent velocity measurements, $\sigma_i$ and $\sigma_j$ are their internal uncertainties (\ie\ 0.85 times the half-width half-max errors reported by {\tt emsao}), and the sum is taken over all observations performed with a similar spectrograph+grating configuration.

\begin{deluxetable*}{crrlcrrlcrrlcrrlcrrl}
\tabletypesize{\scriptsize}
\tablecaption{Mean Velocity Offsets Between Setups\label{tabBetSetups}}
\tablewidth{0pt}
\tablehead{&\multicolumn{3}{c}{IC 342} &&\multicolumn{3}{c}{M74} 
&&\multicolumn{3}{c}{M83} &&\multicolumn{3}{c}{M94} 
&&\multicolumn{3}{c}{M101}\\
Setup &N &$\langle v \rangle$ &$\sigma_{\langle v \rangle}$
&&N &$\langle v \rangle$ &$\sigma_{\langle v \rangle}$
&&N &$\langle v \rangle$ &$\sigma_{\langle v \rangle}$
&&N &$\langle v \rangle$ &$\sigma_{\langle v \rangle}$
&&N &$\langle v \rangle$ &$\sigma_{\langle v \rangle}$
}
\startdata
1 & 16 & 1.4 & 2.7 && 17 & $-1.0$ & 1.5 && 65 & 3.0 & 0.8 && 14 & $-2.4$ & 1.5 && 20 & 6.1 & 2.4 \\
2 & 20 & $-1.0$ & 2.8 && 18 & 4.2 & 1.4 && 44 & 5.3 & 2.0 && 14 & 3.3 & 2.3 && 21 & 1.0 & 2.5 \\
3 & 13 & 1.2 & 3.8 && 11 & 0.8 & 2.4 && 51 & 1.1 & 1.2 && 14 & 2.7 & 1.6 && 22 & $-4.6$ & 4.1 \\
4 & 16 & $-4.3$ & 2.8 && 15 & $-2.6$ & 2.6 && 56 & $-4.1$ & 1.4 && 21 & 0.4 & 1.1 && 24 & 2.1 & 2.4 \\
5 & 20 & 1.1 & 0.9 && 19 & $-3.3$ & 3.1 && 57 & 0.7 & 1.2 && 19 & 0.6 & 0.9 && 15 & 8.2 & 4.0 \\
6 & 14 & 0.7 & 1.6 && 14 & $-2.4$ & 1.5 && 66 & $-3.5$ & 0.8 && 14 & $-1.4$ & 4.0 && 22 & 2.2 & 3.2 \\
7 & 9 & 0.1 & 3.7 && 9 & 3.9 & 4.0 && 50 & $-0.4$ & 0.8 && 13 & $-4.0$ & 1.8 && 18 & 3.4 & 1.8 \\
8 & 8 & $-4.0$ & 3.7 && 18 & $-0.6$ & 1.2 && 52 & 1.7 & 0.6 && 17 & 0.2 & 0.8 && \dots & \dots & \dots \\
9 & 10 & 9.2 & 4.3 && 8 & $-1.4$ & 2.9 && \dots & \dots & \dots && 15 & 4.7 & 2.0 && \dots & \dots & \dots \\
10 & 12 & 2.7 & 2.5 && 13 & $-0.3$ & 1.4 && \dots & \dots & \dots && 15 & 0.1 & 0.7 && \dots & \dots & \dots \\
11 & 14 & 1.9 & 3.4 && 12 & $-4.0$ & 4.3 && \dots & \dots & \dots && 18 & $-0.9$ & 1.1 && \dots & \dots & \dots \\
12 & \dots & \dots & \dots && 15 & 0.3 & 2.6 && \dots & \dots & \dots && \dots & \dots & \dots && \dots & \dots & \dots \\
13 & \dots & \dots & \dots && 6 & $-0.8$ & 4.5 && \dots & \dots & \dots && \dots & \dots & \dots && \dots & \dots & \dots \\
14 & \dots & \dots & \dots && 8 & 0.7 & 2.6 && \dots & \dots & \dots && \dots & \dots & \dots && \dots & \dots & \dots \\
\enddata
\end{deluxetable*}

This statistic confirms that the errors produced by {\tt emsao} do indeed represent our true measurement scatter.   For the Blanco+632@10.8 data in M83, our pairwise comparison of 482 independent PN measurements yields a reduced $\chi^2$ value of 0.96 for 444 degrees of freedom.  This is well within the 90\% confidence range of $0.89 < \chi^2 < 1.11$.  Similarly, our dataset of 670 WIYN+VPH velocities for the PNe of IC~342, M74, and M94 produce $\chi^2 = 0.89$ for 503 degrees of freedom.   This low value, which is just outside the 90\% confidence interval $0.90 < \chi^2 < 1.11$, suggests that, if anything, the uncertainties given by {\tt emsao} are overestimates.  Only in M101, where we combined Hydra and MRS data, did {\tt emsao} underestimate the velocity errors, giving a reduced $\chi^2 = 1.44$.  For these data, we must increase the quoted errors by $\sim$15\% to make the internal and external errors consistent.

\subsection{An External Test in M94}
We can perform one more test of our spectroscopy by taking advantage of existing PN data in M94.  A decade ago, \citet{M94PNS} measured 67 emission-line velocities using a dual-beam, ``counter-dispersed imaging'' spectrograph, designed to create H$\alpha$ images with one arm, and obtain radial velocities via \Oline\ slitless spectroscopy with the other.  These authors surveyed two fields in the galaxy, one along the major axis, and one on the minor axis, and measured 53 and 14 PN candidates, respectively.  Their quoted error on these velocities is $\sim$10~\kms.

In Paper~I, we compared the photometric properties of our PNe to those derived by \citet{M94PNS} and found generally good agreement, with 44 objects common to both datasets.   We now have spectra for 37 of these planetaries: 31 along the major-axis, and 6 from the minor-axis field.  In both cases, the velocity difference between the two samples is close to that expected: for the major axis sample, $\sigma = 10.2$~\kms, while the six objects on the minor axis have $\sigma = 14.6$~\kms.   This again implies that our velocity errors are less than our targeted goal of 15~\kms.  There is a zero point shift between the two fields, with our velocities being systematically lower by $45.9 \pm 1.8$~\kms\ along the major axis, and higher by $45.3 \pm 5.9$~\kms\ near the minor axis.  However, this offset is likely due to a zero-point drift in the \citet{M94PNS} observations, since, as the authors point out, their prototype instrument had flexure problems at the telescope.  This introduced a significant error into their absolute velocity scale, although it did not affect the measurement of relative motions.   Thus, the data provide an additional, independent confirmation of our velocities and velocity uncertainties.  As mentioned above, such knowledge is critical for the kinematic study of Paper~III.

\section{IDENTIFYING CONTAMINANTS}
\subsection{\ion{H}{2} Regions}
\citet{p12} have shown that the ratio of \Oline\ to H$\alpha$ is an excellent tool for discriminating PNe from \ion{H}{2} regions.  When the ratio $R = I(\lambda 5007)_0/I({\rm H}\alpha+$[\ion{N}{2}])$_0$ is plotted against absolute [\ion{O}{3}] magnitude (where the apparent [\ion{O}{3}] magnitude is given by $m_{5007} = -2.5 \log F_{5007} - 13.74)$, true PNe occupy a wedge, which is empirically fit by
\begin{equation}
4 > \log R > -0.37 \, M_{5007} - 1.16.\label{eqsquiggle1}
\end{equation}
Practically speaking, this means that PNe in the top $\sim$1.5~mag of the [\ion{O}{3}] planetary nebula luminosity function always have [\ion{O}{3}] $\lambda 5007$ brighter than H$\alpha$.  This contrasts with the line ratios of the vast majority of \ion{H}{2} regions, which have H$\alpha$ as their dominant emission feature \citep[\eg][]{shaver, kniazev, pena}.

Though the photometric survey of Paper~I has already eliminated most compact \ion{H}{2} regions from our list of PN candidates, it is possible that a few such objects slipped through due to uncertain photometry.  More importantly, errors in the Hydra positioner can cause fibers to miss their intended PNe and instead fall on nearby star-forming regions or supernova remnants.  Finally, because all five of our galaxies have a high star-formation rate, diffuse line emission from interstellar material is ubiquitous and often comparable in brightness to the lines of the target PNe.  Thus each spectrum must be examined, to make sure that the observed line ratios are consistent with those expected from an [\ion{O}{3}]-bright planetary nebula.

To derive these ratios, we needed to obtain an approximate flux calibration for our fiber spectra.  Specifically, we needed to estimate the spectral efficiency around H$\alpha$ relative to that at 5007~\AA\null.   This was done in a number of ways.  For the M83 data, which extends from 4500~\AA\ to 7000~\AA, the process was straightforward:  we used the [\ion{O}{3}] and H$\alpha$+[\ion{N}{2}] photometry of Paper~I to derive the expected response between the red and the blue ($F_{\rm phot}$) in the case of uniform efficiency.  We then examined plots of the observed spectroscopic flux ratio, $F_{\rm spec}$, and the photometric to spectroscopic ratio,  $F_{\rm phot}/F_{\rm spec}$, as a function of $F_{\rm phot}$.  The former plot revealed a clear linear trend plus some outliers; the latter showed that for most objects, $F_{\rm phot}/F_{\rm spec}\sim 2$.  This factor was then applied globally to the spectra.  The outlying objects whose uncorrected spectroscopic [\ion{O}{3}] to H$\alpha$ ratio was more than a factor of three lower than their photometric value were flagged as possible contaminants.

A similar criterion was used for those M101 PNe measured with the WIYN+600@10.1 instrument configuration, except in this case, no quantitative H$\alpha$ photometry was available.  We therefore had to estimate the efficiency of the instrument from archival data.  Five months prior to our observations, 140 of M33's PNe were observed with WIYN using the same grating and instrument configuration as for M101 \citep{M33}.  Like the M83 planetaries, these PNe also have photometric measurements at H$\alpha$ and \Oline.  Consequently, by comparing the photometric and spectroscopic line ratios of PNe in M33, we were able to estimate the response ratio needed to test for contamination in the M101 dataset.  Obviously, this procedure was not as robust as that for M83:  not only did it rely on observations taken during a different observing run, but, unlike the M83 data, the M101 (and M33) PN observations were performed without an atmospheric dispersion corrector.  Nevertheless, since we were using the data only to exclude the most obvious of interlopers, our conservative rejection criterion should still be valid.

For the remaining WIYN observations, our spectra did not extend far enough into the red to record H$\alpha$.  For these objects, we used H$\beta$ as a surrogate.  First, we analyzed Hydra observations of a standard star to estimate the throughput of \Oline\ relative to nearby H$\beta$.  As expected, the data indicated that there was only a slight ($\sim$8\%) decrease in efficiency between these two wavelengths.  This value (which was close to the $\sim$5\% drop expected from the grating's advertised blaze function), was then used to derive the true [\ion{O}{3}]-H$\beta$ ratio of each object. We then scaled our H$\beta$ values to H$\alpha$, using an estimate of the foreground Galactic extinction \citep{sfd98}, a \citet{ccm89} reddening law with $A_V = 3.1$, and an expected Case~B H$\alpha$ to H$\beta$ ratio of 2.86 \citep{brocklehurst}.  Again, objects with $I(\lambda 5007)/I({\rm H}\alpha$) inconsistent with equation~(\ref{eqsquiggle1}) were tagged as possible contaminants.

We note that this last analysis has two limitations.  The first is that it does not account for the contribution of [\ion{N}{2}] $\lambda\lambda 6548,6584$ to the photometrically defined ratio of equation~(\ref{eqsquiggle1}).  For most PNe, this is not a serious omission: an examination of the \citet{M33} sample of M33 PNe shows that [\ion{N}{2}] can safely be neglected in $\sim$85\% of objects.  Moreover, in IC~342, M74, and M94, the stronger [\ion{N}{2}] emission line at $\lambda 6584$ was redshifted onto the tail of our H$\alpha$ interference filter, and was thus suppressed by $\sim$55\%, $\sim$88\% and $\sim$79\%, respectively.  Consequently, even when [\ion{N}{2}] was strong, it was not contributing much flux to our H$\alpha$ photometry.  A more important problem is that by scaling the H$\beta$ flux by 2.86, we are neglecting the contributions of internal galactic and circumstellar extinction, which can greatly increase this value.  We will consider this effect in \S 7; for now, we will accept the fact that our extrapolation may have a systematic error which would cause us to overestimate a PN's [\ion{O}{3}]/H$\alpha$ ratio.

\begin{figure}[t]
\epsscale{1.2}
\plotone{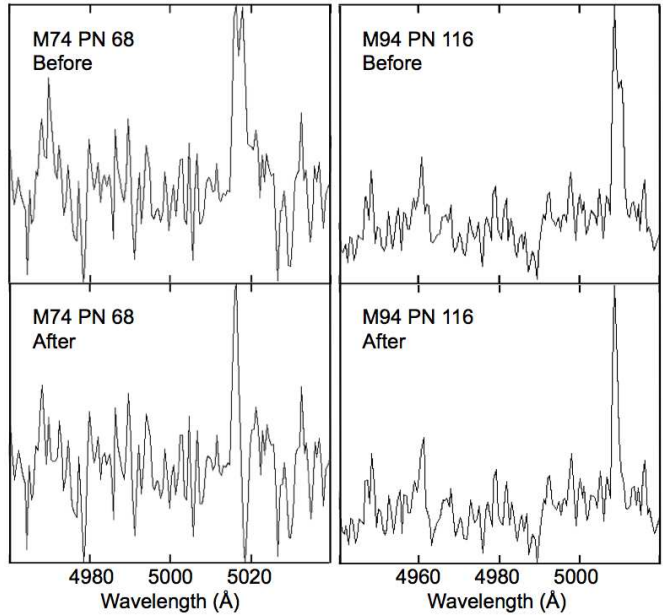}
\caption{Two examples of a PN spectrum corrected for contamination from a low-excitation object.  In both cases, the [\ion{O}{3}] lines are initially double-peaked; after the velocity of H$\beta$ is used to subtract the field emission, the line profiles become consistent with that of the instrument's point spread function. \label{befaft} }
\end{figure}

After examining the PN spectra, we found that $\sim$15\% of our targets had line ratios which suggested contamination by a low-excitation object.  In over half of these cases, the velocity derived from the Balmer lines was indistinguishable from that found from \Oline.  However, in 42 out of the 97 objects, the lines were kinematically different, suggesting flux from two different sources.  In these spectra, \Oline\ was either double-peaked, or had a profile significantly broader than the spectral point-spread-function.  When this occurred, we attempted to subtract off the low-excitation component using the velocity of H$\beta$ (or H$\alpha$) as a guide.  In 25 cases, we were successful in isolating the PN's emission from that of the contaminating source, and could return the object to the kinematic sample.  (See Figure~\ref{befaft} for two examples of these subtractions.)  In the other 17 cases, (which involved the lower resolution M83 and M101 data), we noted the blend, but could not deconvolve the velocities.

Finally, after identifying the possible contaminants, we revisited their positions on the \Oline\ and H$\alpha$ images of Paper~I\null.  For most of the objects, the targeted PN did have a bright, low-excitation source nearby; in these cases, we simply excluded the source from our kinematic sample.  In a few cases where the PN was clearly isolated, we re-examined the line fluxes and line profiles to determine the source of the discrepancy. If we concluded that the recorded spectrum could, indeed, have come from the planetary, we re-classified the object as a PN, and included it in our analysis.

\begin{figure}[b]
\epsscale{1.2}
\plotone{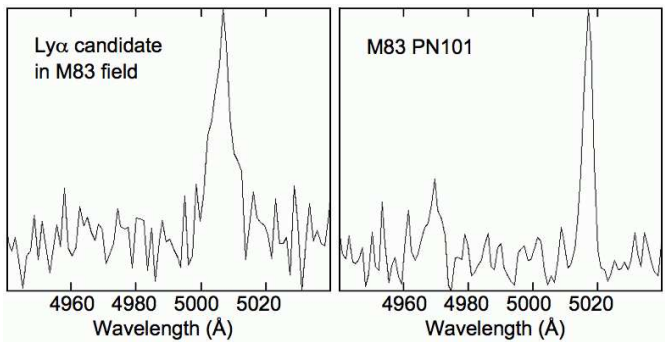}
\caption{Spectrum of a $z = 3.12$ Ly$\alpha$ emitting galaxy in the field of M83, compared to that of a normal planetary of similar brightness.  Note the absence of [\ion{O}{3}] $\lambda 4959$ and the broad, asymmetric line profile.  No other lines are present in the spectrum.  The object is among the 
brightest Ly$\alpha$ galaxies yet discovered. \label{Highz} }
\end{figure}

\subsection{High Redshift Galaxies}
A second possible source of contamination is high redshift galaxies.  At $z \sim 3.12$, starbursting galaxies have their Ly$\alpha$ emission redshifted into the bandpass of our [\ion{O}{3}] filter, and because their observer-frame equivalent widths can be exceedingly large, these objects can easily be confused with planetaries \citep{arnaboldi02, ipn3}.  The limited depth of our survey prevents us from detecting many of these objects:  according to \citet{gronwall}, the luminosity function of $z \sim 3.1$ Ly$\alpha$ emitters (LAEs) in the emission line takes the form of a \citet{schechter} function with $m_{5007}^* \sim 26.9$.  Since this cutoff is well below the limiting magnitude of our photometric surveys (see Paper~I), we would not expect to find many LAEs in our sample.

\begin{figure*}
\epsscale{1.1}
\plottwo{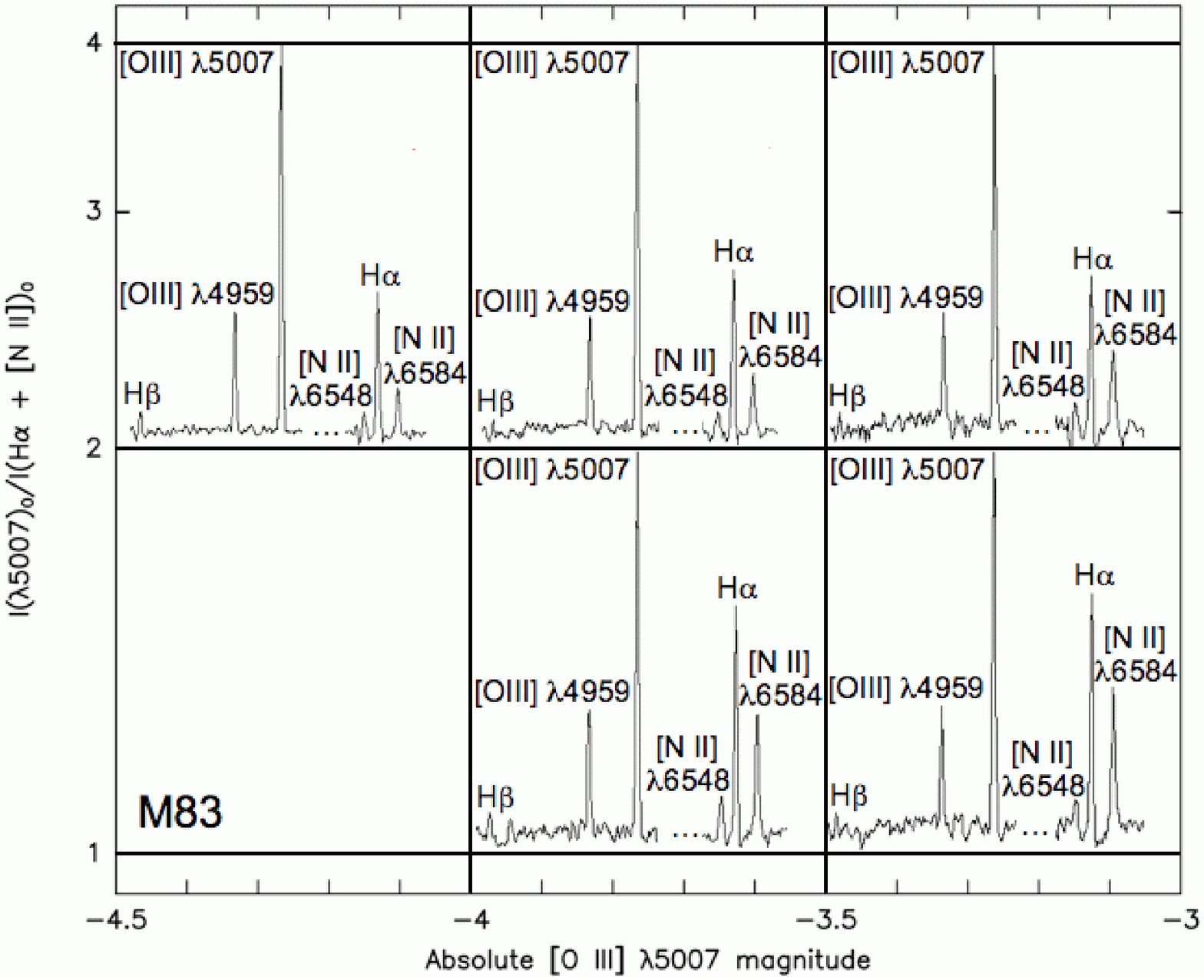}{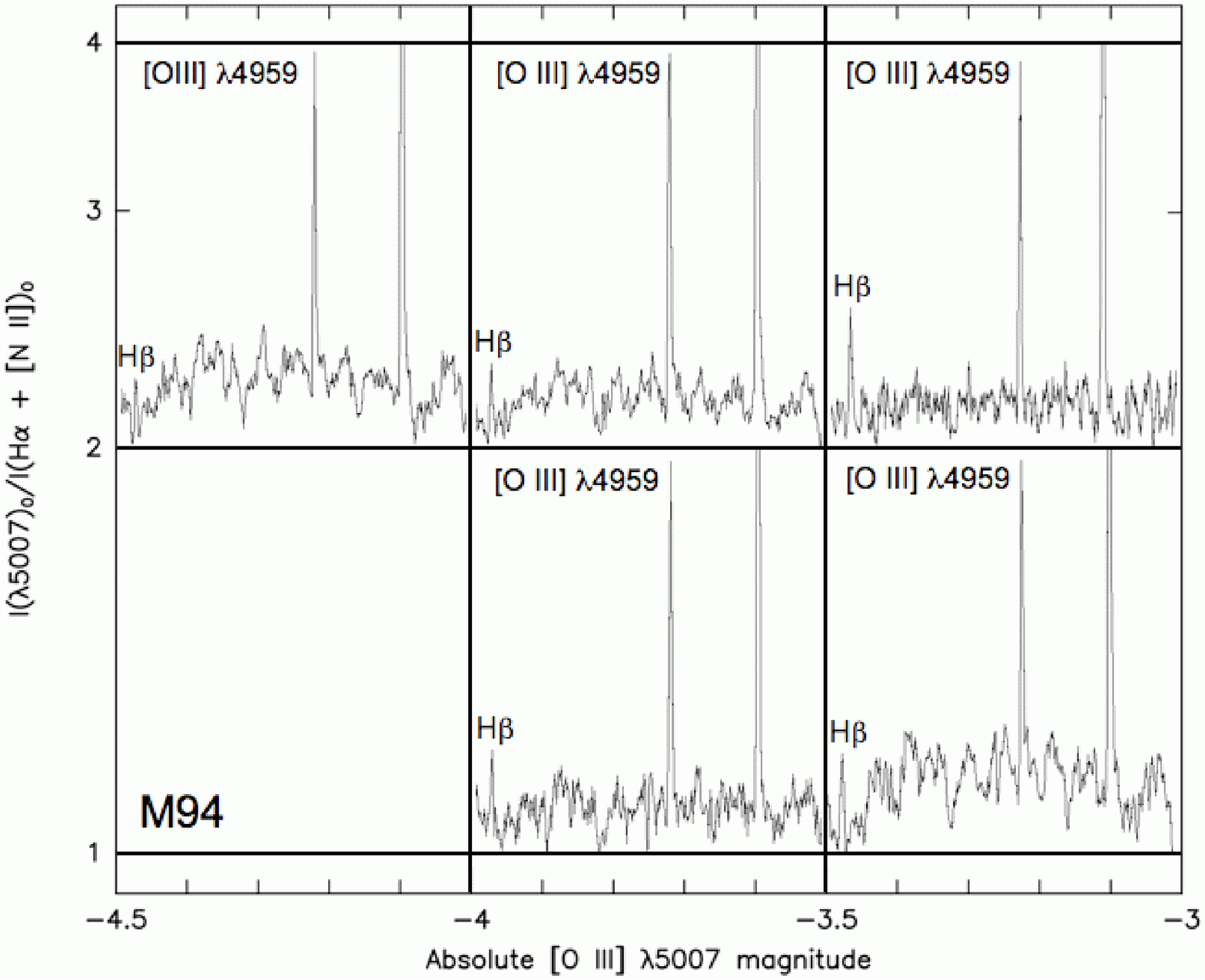}
\caption{Co-added spectra of PNe in M83 and M94, as a function of [\ion{O}{3}] absolute magnitude and [\ion{O}{3}] $\lambda 5007$/H$\alpha$+[\ion{N}{2}] flux ratio.  The abscissa plots wavelength; the ordinate shows relative counts, with the red portion of the M83 spectra reduced by a factor of two to correct for the higher system throughput.  The spectra with $I(\lambda5007)_0/I($H$\alpha$+[\ion{N}{2}])$_0 > 2$ have been co-added in the upper rows, while those with a flux ratio $< 2$ have been co-added in the lower rows.  The spectra are remarkably similar:  the only obvious change is the increased importance of [\ion{N}{2}] in lower luminosity objects. \label{CoAdd} }
\end{figure*}

Still, it is possible for some extremely bright LAEs to masquerade as PNe.  If we extrapolate the \citet{schechter} function of \citet{gronwall} to the limiting magnitude of each survey, then we should expect to find $\sim$0.2~LAEs in the field of M101, $\sim$0.5~LAEs near M74, and $\sim$0.8~LAEs in the wide-field Mosaic frame covering M83.  (LAEs in the fields of M94 and IC~342 are exceedingly unlikely, $\sim$2$\times 10^{-3}$ and 4$\times 10^{-7}$, respectively).  In fact, one LAE was detected in our survey.  The object is in the field of M83 ($\alpha(2000) = 13^h38^m09.26^s$, $\delta(2000)$ = $-29^{\circ}38\arcmin31.5\arcsec$) and is easily recognizable via its strong, asymmetric line profile (centered at 5006.8~\AA), its non-negligible velocity width ($\sim$400~\kms\ full-width-half-maximum), and an absence of any accompanying emission at the wavelengths of [\ion{O}{3}] $\lambda 4959$ or H$\beta$.  The object cannot be a background source at $z \sim 0.34$, since redshifted H$\beta$ is not present at $\sim$6530~\AA, and the object's photometrically inferred equivalent width ($\sim$600~\AA) is much larger than that of a typical [\ion{O}{2}] galaxy \citep{hogg98}.  Its line profile and line width are, however, consistent with those observed for other high-redshift Ly$\alpha$ emitters \citep[\eg][]{dawson}. 

Interestingly, the bright apparent magnitude of our Ly$\alpha$ emitter ($m_{5007} \sim 25.15$, or $\log F_{5007} = -15.556$) makes it one of the most luminous $z \sim 3.1$ LAEs ever observed, with a total emission-line luminosity of $\sim$2.3$\times 10^{43}$~ergs~s$^{-1}$.  Most similarly bright LAEs harbor an AGN \citep{gronwall, ouchi}, but in this object, C~IV $\lambda 1550$ is not seen, nor is there any evidence for a broad line component to Ly$\alpha$.  If the source is indeed powered solely by star-formation, then the Case B-based relations of \citet{kennicutt} and \citet{hu} imply a total star-formation rate of $\gtrsim 20 \, M_{\odot}$~yr$^{-1}$.  This value is larger than any of the rates derived by \citet{gronwall} for a sample of LAEs in the Extended Chandra Deep Field South, and is comparable to that associated with the brightest $z \sim 3.1$ LAEs identified by \citet{ouchi}.  Figure~\ref{Highz} displays the emission-line profile of this object.

\section{FINAL PN VELOCITIES}
In measuring our velocities, we have relied solely on the wavelength of the extremely strong [\ion{O}{3}] $\lambda 5007$ emission line and have ignored the information contained in the weaker lines of [\ion{O}{3}] $\lambda 4959$, H$\alpha$, H$\beta$,  and [\ion{N}{2}] $\lambda\lambda 6548,6584$.  With our final PN velocities secured, we could check this decision.  To do this, we began by selecting all those PNe with non-contaminated, well-measured ($\sigma_v < 15$~\kms) emission lines.  We then shifted these spectra into the rest frame, and co-added the data, to create a single high signal-to-noise template for each telescope+spectrograph configuration.  This template was then cross-correlated against the individual PN spectra using the {\tt xcsao} task of {\tt rvsao} to derive an alternate measure of velocity. 

\begin{deluxetable*}{lcccrccccrcl}
\tabletypesize{\scriptsize}
\tablecaption{Planetary Nebula Identifications\label{tabPNe}}
\tablewidth{0pt}
\tablehead{&&&&&&&&&\colhead{$v_{\odot}$} &\colhead{$\sigma_v$} & \\
\colhead{ID} &\colhead{$\alpha(2000)$} &\colhead{$\delta(2000)$} 
&\colhead{$m_{5007}$} &\colhead{$R$\tablenotemark{a}} &\colhead{$\sigma_{R}$} 
&\colhead{Type\tablenotemark{b}} &\colhead{N$_{\rm trg}$\tablenotemark{c}} 
&\colhead{N$_{\rm det}$\tablenotemark{d}} &\colhead{(\kms)} &\colhead{(\kms)} 
&\colhead{Notes}
}
\startdata
IC 342-1  &03:46:54.99  &+68:06:39.1  &25.26  &3.18  &0.76  &Phot  &3  &3  &73.7  &4.3  & \\
IC 342-165  &03:47:04.81  &+68:04:31.6  &27.67  &0.68  &0.35  &Phot  &3  &0 &\dots  &\dots  & \\
M74-1  &01:36:39.95  &+15:47:02.5  &25.44  &3.09  &\dots  &Spec  &4  &4  &657.3 &4.2  & \\
M74-153  &01:36:57.04  &+15:46:48.2  &27.55  &$>0.52$  &\dots  &Phot  &0  &0  &\dots  &\dots  & \\
M83-1  &13:37:00.99  &$-$29:57:31.8  &24.24  &2.39  &0.23  &Phot  &2  &2  &586.9  &2.5  & \\
M83-241  &13:37:21.58  &$-$29:58:37.3  &27.15  &$>0.60$  &\dots  &Phot  &0  &0  &\dots  &\dots  & \\
M94-1  &12:50:52.17  &+41:08:36.2  &23.86  &2.78  &0.16  &Phot  &1  &1  &317.4 &2.0  &\tablenotemark{D} \\
M94-150  &12:50:52.03  &+41:11:22.5  &26.21  &0.99  &0.38  &Phot  &3  &3  &377.4  &5.8  &\tablenotemark{D} \\
M101-1  &14:02:40.57  &+54:13:56.3  &24.97  &\dots  &\dots  &\dots  &4  &4  &140.7  &4.2  &\tablenotemark{C} \\
M101-65  &14:03:05.94  &+54:25:37.2  &26.43  &\dots  &\dots  &\dots  &2  &1  &221.9  &12.6  & \\
\enddata
\tablecomments{C: Emission line is a blend of multiple components, dominated by a low-excitation contaminant; D: PN first measured by \citet{M94PNS}; H: Spectrum likely that of a nearby \ion{H}{2} region; P: Emission line is a blend of multiple components, dominated by the planetary; S: Emission line is a blend, with the velocity obtained by subtracting off the low excitation component; U: Not part of our analysis, with $\sigma_v > $ 15 \kms; X: Velocity and uncertainty derived from {\tt xcsao}.  Table 4 is published in its entirety in the electronic edition of the {\it Astrophysical Journal}.  A portion is shown here for guidance regarding its form and content.}
\tablenotetext{a}{$R = I(\lambda 5007)_0/I({\rm H}\alpha+$[\ion{N}{2}])$_0$}
\tablenotetext{b}{Type of $R$ value given}
\tablenotetext{c}{Number of Hydra+MRS setups in which PN was targeted}
\tablenotetext{d}{Number of Hydra+MRS setups in which PN was detected}
\end{deluxetable*}

\begin{deluxetable*}{lcccccccc}
\tabletypesize{\scriptsize}
\tablecaption{Number of PNe Targeted\label{tabTargDet}}
\tablewidth{0pt}
\tablehead{&\colhead{Photometric} &\colhead{Not} &\colhead{Not} 
&\colhead{Probable} &\multicolumn{3}{c}{Number of Detections} 
&\colhead{Total} \\
\colhead{Galaxy} &\colhead{Sample} &\colhead{Targeted} &\colhead{Detected}
&\colhead{Contaminant} &\colhead{$N=1$} &\colhead{$N=2$} &\colhead{$N>2$} 
&\colhead{$\sigma_v <$ 15 \kms}
}
\startdata
IC~342 & 165 & 29 & 30 &  3 & 48 & 32 & 23 &  99 \\
M74    & 153 & 28 & 13 &  7 & 43 & 39 & 23 & 102 \\
M83    & 241 & 25 & 12 & 31 & 36 & 66 & 71 & 162 \\
M94    & 150 & 13 &  7 &  2 & 68 & 32 & 28 & 127 \\
M101   & 65  &  1 &  1 & 12 & 13 & 13 & 25 &  48 \\
Total  & 774 & 96 & 63 & 55 &208 &182 &170 & 538 \\
\enddata
\end{deluxetable*}

Figure~\ref{CoAdd} illustrates the validity of this approach.  In the figure, we display a sample of co-added spectra using the Blanco+632@10.8 observations in M83 and the WIYN+VPH data in M94.  To illustrate the consistency of the spectra, both datasets have been corrected for Galactic extinction (using the \citet{ccm89} reddening law) and broken down by absolute [\ion{O}{3}] $\lambda 5007$ magnitude (using 0.5~mag bins and the galactic distances of Paper~I) and excitation (with [\ion{O}{3}]/H$\alpha$ $R = 2$ as the dividing line).  As expected, the weaker Balmer and [\ion{N}{2}] lines are no wider than those of [\ion{O}{3}], demonstrating that our {\tt emsao} velocities are reasonably precise.   More importantly, the mean PN spectra appear remarkably similar from one magnitude range to the next.  In M83, the only observable change involves the strength of [\ion{N}{2}], which increases in importance as one proceeds down the luminosity function.  In the higher resolution M94 data, the strength of H$\beta$ appears anomalously bright in the faintest, high-excitation bin, but this may simply be a stochastic result.  Overall, the principal bright lines of our observed PNe are remarkably consistent across the entire range of our data, justifying the use of a single template spectrum for our analysis.

A comparison of the results reveals that while the {\tt emsao} and {\tt xcsao} velocities always agree, the formal errors produced by the first package are almost always smaller than those quoted by the latter.  Rather than improving our velocities, it appears that by including the spectral regions surrounding the weaker lines, we add in more noise than signal.  Since we know from our analysis of variance (see section \S 4) that the uncertainties quoted by {\tt emsao} are accurate, the use of the simpler routine is fully justified.

Table~\ref{tabPNe} gives our final list of PNe, including their positions, \Oline\ magnitudes, values of $R$ (corrected for foreground Galactic extinction), velocities, and velocity uncertainties.  Unless otherwise noted, these velocities and their errors come from {\tt emsao}.  Out of the 774~PNe detected photometrically, 70\% were measured to a radial velocity accuracy better than 15 \kms.  Of the remaining objects, 96 were not targeted for spectroscopy, either due to fiber-positioning constraints or the apparent faintness of the \Oline\ line.   Only 63 PNe were observed but not detected, with most of these having magnitudes well down the planetary nebula luminosity function.  Table~\ref{tabTargDet} summarizes our results.  Tables~\ref{tabN5068} and \ref{tabN6946} give the positions and photometric properties of PNe in galaxies which were observed in Paper~I, but not selected for spectroscopic follow-up.

\section{PNe AND CIRCUMSTELLAR EXTINCTION}
In Paper~III, we will use our planetary nebula velocities to probe the velocity dispersion and disk mass distribution of galactic disks.  Ideally, it would be helpful to couple this kinematic data with detailed information about the population of PN progenitors.  Unfortunately, this is not easy to do.  Just because stars with initial masses between $\sim$1$M_{\odot}$ and $\sim$5$M_{\odot}$ are expected to evolve into planetary nebulae, that does not mean that all intermediate mass stars contribute equally to the bright end of the planetary nebula luminosity function.  If the models by \citet{marigo} and \citet{mendez} are correct, then the bright-end of the planetary nebula luminosity function is dominated by objects from relatively massive ($\sim$2$M_{\odot}$), relatively young ($\sim$1~Gyr) progenitors.  On the other hand, a number of recent analyses suggest that most [\ion{O}{3}]-bright planetaries are not formed from single stars at all.  Alternate scenarios, involving common-envelope interactions \citep{frank}, blue straggler evolution \citep{bs-pn}, and even ionization from accreting white dwarfs \citep{soker} have all been used to explain the PN phenomenon.   Thus, at present, it is impossible to say anything definitive about the progenitors of our kinematic test particles.

We can, however, combine our spectroscopic line ratios with the [\ion{O}{3}] $\lambda 5007$ and H$\alpha$+[\ion{N}{2}] photometry of Paper~I to gain some insight into the question of the PN population's homogeneity.  Specifically, we can use our data to estimate circumstellar extinction, and probe the uniformity of the PN population as a function of absolute magnitude.  A PN whose central star is intrinsically faint can have two possible progenitors:  it can be a high core-mass, faded remnant that was once at the bright-end cutoff of the PN luminosity function, or it can be a lower-mass star which is just now attaining maximum brightness.   In theory, these two scenarios make different predictions about the behavior of the AGB dust envelope.  In the case of a high-mass star, the circumstellar extinction should remain roughly constant, since the timescale for central star evolution is much shorter than that for nebular expansion.  However, for lower-mass stars, the slower evolutionary timescales allow dust to dissipate and produce systematically less extinction.  By measuring the Balmer decrements in a sample of planetary nebulae of varying brightnesses, it may be possible to perform a global test of the PN population.  Note that this type of analysis can only be performed on a sample of extragalactic PNe, since in the Galaxy, the uncertainties associated with distance and foreground reddening overwhelm all other aspects of the analysis.

The key to performing this experiment is to have well-determined estimates of H$\beta$ relative to [\ion{O}{3}] $\lambda 5007$.  As Fig.~\ref{CoAdd} illustrates, such data are not easy to obtain:  in our sample of galaxies, only M94 has a sufficient number of high quality measurements.   To this sample, we can add in the PNe of M33, which have photometry and spectroscopy from \citet{M33}, and the LMC, where spectroscopic and photometric data are available from \citet{md1, md2}, \citet{vdm} and \citet{p6}.   We then estimate each PN's Balmer decrement:  for M94 PNe, we simply measure the ratio of H$\beta$ to [\ion{O}{3}] $\lambda 5007$ on our spectra, and then scale these values to H$\alpha$ using our absolute [\ion{O}{3}] and H$\alpha$+[\ion{N}{2}] photometry.  Such a procedure neglects the contribution of the nitrogen lines, but since the stronger of these lines, [\ion{N}{2}] $\lambda 6584$, falls on the wings of the narrow-band filter's bandpass (where the transmission is $\sim$1/4 maximum), this is not a serious problem.  In the case of M33 and the LMC, where the spectral region about H$\alpha$ is directly observed, we mimic our M94 data by including [\ion{N}{2}] $\lambda 6548$ and one quarter of [\ion{N}{2}] $\lambda 6584$ in our estimate of H$\alpha$.  We then compute the logarithmic H$\beta$ extinction for each planetary, using our faux H$\alpha$/H$\beta$ ratio, an assumed intrinsic decrement of 2.86, and a \citet{ccm89} reddening law. 

\begin{figure}[t]
\epsscale{1.05}
\plotone{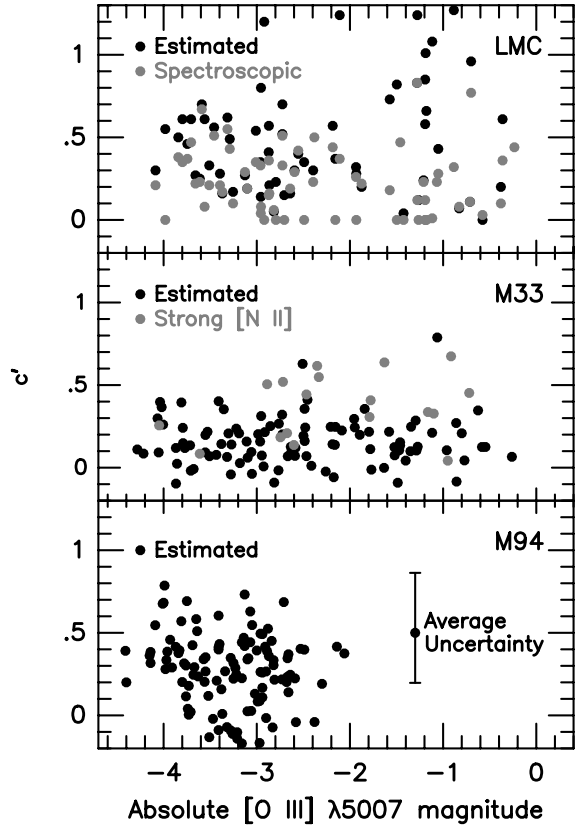}
\caption{Balmer-decrement based estimated values, $c^{\prime}$, for the logarithmic extinction at H$\beta$ for PNe in the LMC, M33, and M94, as a function of [\ion{O}{3}] $\lambda 5007$ absolute magnitude.  These values are slightly overestimated, since H$\alpha$ is partially contaminated by emission from [\ion{N}{2}].  Spectroscopically derived values of $c$ (without the [\ion{N}{2}] adjustment) in the Magellanic Cloud are shown as gray circles.  Note the general agreement between galaxies; there is no evidence for an extinction dependence on metallicity, nor is there any significant trend with absolute magnitude.  \label{cval} }
\end{figure}

The results of this analysis are shown in Figure~\ref{cval}.  From the figure, it is clear that the mean circumstellar extinction for PNe in M94 is roughly the same as that observed in M33 and the LMC, \ie\ $E(B-V) \sim 0.2$.  This is a somewhat surprising result:  one might reasonably expect the PNe formed from the metal-poor stars of the LMC to have systematically less dust.  This does not appear to be the case.  More importantly, the data show no evidence for a gradient in extinction with absolute [\ion{O}{3}] magnitude.  Note that, as one goes down the luminosity function, [\ion{N}{2}] increases in importance.  So, if nitrogen emission were contaminating our analysis, we would expect to see an inverse correlation between estimated extinction and absolute magnitude.  But the data show no significant trend at all; this argues against any large change in stellar population with absolute [\ion{O}{3}] magnitude.  Formally, our results are consistent with most of the PNe in our sample beginning as high core-mass stars before fading.  However, considering the scatter in the data and the uncertainties, our results only weakly support this conclusion.  Small population differences could also be consistent with our results.

\section{CONCLUSIONS}
In Paper~I and in \citet{M101PNe} we identified 774 planetary nebula candidates in five nearby, low-inclination galaxies: IC~342, M74, M83, M94, and M101.  We now have spectroscopic confirmations for over 600 of the PNe, with 538 having been measured to better than 15~\kms\ precision.  Of the remaining $\sim$160~objects, most were either not targeted for spectroscopy (due to fiber positioning constraints) or too faint to be detected with our spectrographs.   We can now use these data to investigate the kinematic structure of the galaxies' disks and, for the first time, dynamically measure disk mass-to-light ratios throughout the body of large spiral galaxies.

\begin{deluxetable}{lcccrc}
\tabletypesize{\scriptsize}
\tablecaption{NGC~5068 Planetary Nebula Candidates\label{tabN5068}}
\tablewidth{0pt}
\tablehead{
\colhead{ID} &\colhead{$\alpha(2000)$} &\colhead{$\delta(2000)$}
&\colhead{$m_{5007}$} &\colhead{$R$} &\colhead{$\sigma_{R}$}
}
\startdata
NGC 5068-1  &13:18:59.79  &$-2$1:00:07.5  &24.85  &2.32  & 0.21 \\
NGC 5068-2  &13:19:03.17  &$-2$1:01:59.3  &24.94  &3.76  & 0.61 \\
NGC 5068-3  &13:18:51.10  &$-2$1:04:24.9  &24.97  &2.29  & 0.21 \\
NGC 5068-4  &13:18:47.30  &$-2$1:02:49.7  &25.04  &1.70  & 0.16 \\
NGC 5068-5  &13:19:01.06  &$-2$1:00:49.8  &25.07  &1.61  & 0.11 \\
NGC 5068-6  &13:19:02.99  &$-2$1:02:10.8  &25.13  &1.87  & 0.24 \\
NGC 5068-7  &13:18:48.73  &$-2$1:00:09.3  &25.28  &2.29  & 0.33 \\
NGC 5068-8  &13:19:11.11  &$-2$0:59:57.5  &25.30  &3.18  & 0.39 \\
NGC 5068-9  &13:18:57.66  &$-2$1:05:44.5  &25.32  &1.76  & 0.20 \\
NGC 5068-10 &13:19:01.92  &$-2$1:01:35.9  &25.40  &1.65 & 0.17 \\
NGC 5068-11 &13:18:49.66  &$-2$1:04:20.1  &25.40  &$>1.69$ & \dots \\
NGC 5068-12 &13:18:53.90  &$-2$1:00:56.9  &25.41  &1.80  & 0.43 \\
NGC 5068-13 &13:18:53.41  &$-2$0:59:59.2  &25.49  &1.66  & 0.25 \\
NGC 5068-14 &13:18:44.62  &$-2$1:00:01.8  &25.50  &1.54  & 0.15 \\
NGC 5068-15 &13:18:47.50  &$-2$1:01:01.2  &25.57  &1.18  & 0.10 \\
NGC 5068-16 &13:19:01.00  &$-2$1:03:40.2  &25.96  &1.26  & 0.21 \\
NGC 5068-17 &13:18:53.33  &$-2$0:59:51.9  &25.98  &0.83  & 0.11 \\
NGC 5068-18 &13:18:42.89  &$-2$1:03:04.2  &26.02  &0.91  & 0.16 \\
NGC 5068-19 &13:18:53.26  &$-$20:59:18.3  &26.11  &1.07  & 0.16 \\
\enddata
\end{deluxetable}

\acknowledgments
We would like to thank KPNO and CTIO personnel for friendly travel, telescope, and instrumental support (especially Di Harmer for her excellent assistance with WIYN+Hydra) and the HET RAs.  We would also like to thank John Feldmeier, George Jacoby, Antonino Cucchiara, Matt Vinciguerra, Patrick Durrell, and Kenneth Moody for help with observing and would like to acknowledge the useful comments of our anonymous referee.  This research has made use of the USNOFS Image and Catalogue Archive operated by the United States Naval Observatory, Flagstaff Station (http://www.nofs.navy.mil/data/fchpix/), NASA's Astrophysics Data System, and the NASA/IPAC Extragalactic Database (NED) which is operated by the Jet Propulsion Laboratory, California Institute of Technology, under contract with the National Aeronautics and Space Administration.  This work was supported by NSF grant AST 06-07416 and a Pennsylvania Space Grant Fellowship.

Facilities: \facility{Blanco(Hydra)}, \facility{WIYN(Hydra)}, \facility{HET(MRS)}.

\begin{deluxetable}{lcccc}
\tabletypesize{\scriptsize}
\tablecaption{NGC~6946 Planetary Nebula Candidates\label{tabN6946}}
\tablewidth{0pt}
\tablehead{
\colhead{ID} &\colhead{$\alpha(2000)$} &\colhead{$\delta(2000)$}
&\colhead{$m_{5007}$} &\colhead{$R$} 
}
\startdata
NGC 6946-1   &20:34:39.26  &+60:04:44.7  &25.61  &$>0.90  $ \\
NGC 6946-2   &20:35:25.42  &+60:08:37.6  &25.72  &$>1.16  $ \\
NGC 6946-3   &20:34:57.58  &+60:08:11.4  &25.72  &$>1.01  $ \\
NGC 6946-4   &20:34:26.31  &+60:07:38.2  &25.76  &$>1.76  $ \\
NGC 6946-5   &20:35:18.35  &+60:07:29.4  &25.79  &$>0.91  $ \\
NGC 6946-6   &20:34:21.94  &+60:07:30.1  &25.79  &$>0.95  $ \\
NGC 6946-7   &20:34:56.08  &+60:06:36.4  &25.80  &$>0.91  $ \\
NGC 6946-8   &20:34:40.98  &+60:04:41.5  &25.81  &$>0.90  $ \\
NGC 6946-9   &20:35:05.18  &+60:12:26.9  &25.87  &$>2.55  $ \\
NGC 6946-10  &20:34:33.63  &+60:05:22.6  &25.89  &$>2.75  $ \\
NGC 6946-11  &20:34:21.69  &+60:11:47.3  &25.91  &$>1.44  $ \\
NGC 6946-12  &20:35:06.80  &+60:10:14.7  &25.95  &$>1.20  $ \\
NGC 6946-13  &20:35:33.83  &+60:11:53.1  &25.96  &$>0.86  $ \\
NGC 6946-14  &20:34:35.39  &+60:05:00.5  &25.96  &$>1.35  $ \\
NGC 6946-15  &20:35:06.52  &+60:12:49.5  &25.98  &$>1.66  $ \\
NGC 6946-16  &20:34:38.62  &+60:06:31.9  &25.99  &$>0.73  $ \\
NGC 6946-17  &20:34:24.38  &+60:06:16.9  &26.01  &$>0.74  $ \\
NGC 6946-18  &20:34:43.52  &+60:07:51.9  &26.01  &$>1.58  $ \\
NGC 6946-19  &20:34:31.10  &+60:05:59.7  &26.03  &$>0.72  $ \\
NGC 6946-20  &20:34:17.78  &+60:09:42.2  &26.04  &$>1.04  $ \\
NGC 6946-21  &20:35:00.52  &+60:08:58.5  &26.05  &$>0.68  $ \\
NGC 6946-22  &20:35:01.74  &+60:07:31.3  &26.11  &$>1.33  $ \\
NGC 6946-23  &20:34:38.03  &+60:07:33.3  &26.11  &$>1.28  $ \\
NGC 6946-24  &20:34:37.65  &+60:11:56.4  &26.11  &$>0.94  $ \\
NGC 6946-25  &20:35:13.96  &+60:11:37.0  &26.12  &$>1.12  $ \\
NGC 6946-26  &20:34:33.41  &+60:10:41.2  &26.14  &$>1.01  $ \\
NGC 6946-27  &20:35:21.04  &+60:04:42.8  &26.15  &$>1.17  $ \\
NGC 6946-28  &20:35:18.84  &+60:11:40.2  &26.15  &$>1.12  $ \\
NGC 6946-29  &20:35:20.54  &+60:13:56.0  &26.15  &$>0.72  $ \\
NGC 6946-30  &20:35:20.18  &+60:05:05.1  &26.16  &$>0.71  $ \\
NGC 6946-31  &20:34:17.43  &+60:10:48.8  &26.17  &$>1.02  $ \\
NGC 6946-32  &20:34:57.31  &+60:07:16.2  &26.18  &$>0.66  $ \\
NGC 6946-33  &20:34:47.16  &+60:09:06.1  &26.19  &$>1.00  $ \\
NGC 6946-34  &20:34:37.06  &+60:10:26.0  &26.21  &$>0.65  $ \\
NGC 6946-35  &20:34:27.87  &+60:06:11.1  &26.21  &$>1.04  $ \\
NGC 6946-36  &20:35:18.87  &+60:07:17.8  &26.22  &$>1.20  $ \\
NGC 6946-37  &20:34:54.98  &+60:14:03.3  &26.23  &$>1.74  $ \\
NGC 6946-38  &20:34:32.68  &+60:10:42.3  &26.25  &$>0.48  $ \\
NGC 6946-39  &20:34:30.10  &+60:07:36.5  &26.25  &$>0.63  $ \\
NGC 6946-40  &20:34:47.17  &+60:08:25.8  &26.26  &$>0.56  $ \\
NGC 6946-41  &20:35:30.19  &+60:10:44.5  &26.29  &$>1.14  $ \\
NGC 6946-42  &20:35:07.16  &+60:05:33.0  &26.29  &$>0.69  $ \\
NGC 6946-43  &20:34:27.10  &+60:09:49.0  &26.31  &$>1.43  $ \\
NGC 6946-44  &20:34:44.20  &+60:09:38.5  &26.33  &$>0.76  $ \\
NGC 6946-45  &20:35:28.44  &+60:04:48.0  &26.37  &$>1.59  $ \\
NGC 6946-46  &20:34:37.18  &+60:06:26.4  &26.37  &$>0.85  $ \\
NGC 6946-47  &20:35:23.43  &+60:10:24.9  &26.38  &$>0.70  $ \\
NGC 6946-48  &20:35:26.51  &+60:13:45.2  &26.40  &$>1.80  $ \\
NGC 6946-49  &20:35:11.05  &+60:09:58.0  &26.44  &$>2.10  $ \\
NGC 6946-50  &20:35:22.40  &+60:10:56.1  &26.47  &$>0.56  $ \\
NGC 6946-51  &20:35:19.32  &+60:10:29.3  &26.50  &$>0.55  $ \\
NGC 6946-52  &20:35:12.56  &+60:11:19.3  &26.52  &$>0.88  $ \\
NGC 6946-53  &20:35:11.96  &+60:12:00.8  &26.53  &$>1.45  $ \\
NGC 6946-54  &20:35:08.93  &+60:08:29.6  &26.53  &$>0.48  $ \\
NGC 6946-55  &20:35:29.14  &+60:10:09.4  &26.55  &$>1.06  $ \\
NGC 6946-56  &20:35:22.54  &+60:08:11.5  &26.56  &$>0.70  $ \\
NGC 6946-57  &20:34:22.88  &+60:10:14.4  &26.62  &$>0.80  $ \\
NGC 6946-58  &20:34:19.04  &+60:05:15.2  &26.63  &$>0.73  $ \\
NGC 6946-59  &20:34:45.99  &+60:04:30.4  &26.65  &$>0.72  $ \\
NGC 6946-60  &20:34:44.63  &+60:12:53.4  &26.68  &$>1.42  $ \\
NGC 6946-61  &20:35:07.59  &+60:07:45.8  &26.69  &$>0.42  $ \\
NGC 6946-62  &20:34:54.67  &+60:06:09.8  &26.70  &$>0.42  $ \\
NGC 6946-63  &20:34:59.01  &+60:13:37.9  &26.72  &$>0.42  $ \\
NGC 6946-64  &20:34:59.77  &+60:12:26.5  &26.74  &$>0.44  $ \\
NGC 6946-65  &20:35:03.43  &+60:12:49.5  &26.83  &$>0.94  $ \\
NGC 6946-66  &20:35:22.44  &+60:06:53.9  &26.85  &$>0.54  $ \\
NGC 6946-67  &20:34:58.04  &+60:12:59.9  &26.91  &$>0.34  $ \\
NGC 6946-68  &20:35:09.01  &+60:11:42.4  &27.00  &$>0.76  $ \\
NGC 6946-69  &20:34:50.50  &+60:06:54.4  &27.05  &$>0.33  $ \\
NGC 6946-70  &20:34:29.19  &+60:12:26.5  &27.05  &$>0.34  $ \\
NGC 6946-71  &20:34:33.01  &+60:11:36.3  &27.32  &$>0.26  $ \\
\enddata
\end{deluxetable}

\end{document}